\documentclass[10pt,conference]{IEEEtran}
\IEEEoverridecommandlockouts

\usepackage{adjustbox}
\usepackage{xspace}
\usepackage{tabularx}
\usepackage{colortbl}
\usepackage{subfig}
\usepackage{multirow}
\usepackage{booktabs}
\usepackage{listings}
\usepackage{scratch3}
\usepackage{arydshln}
\usepackage{amsmath,mathabx}
\usepackage[capitalise]{cleveref}
\usepackage{cite}
\usepackage{flushend}

\usepackage{paralist}

\usepackage{collcell}

\setscratch{scale=.60}

\newcommand\definetool[2]{\newcommand{#1}{{\textsc{#2}}\xspace}}
\definetool{\Scratch}{Scratch}
\definetool{\leila}{LeILa}
\definetool{\whisker}{Whisker}
\definetool{\litterbox}{LitterBox}
\definetool{\bastet}{Bastet}
\definetool{\scratchblocks}{scratchblocks}

\colorlet{punct}{red!60!black}
\definecolor{background}{HTML}{EEEEEE}
\definecolor{delim}{RGB}{20,105,176}
\colorlet{numb}{magenta!60!black}

\lstdefinelanguage{json}{
    basicstyle=\normalfont\ttfamily,
    numbers=left,
    numberstyle=\scriptsize,
    stepnumber=1,
    numbersep=8pt,
    showstringspaces=false,
    breaklines=true,
    frame=lines,
    backgroundcolor=\color{background},
    literate=
     *{0}{{{\color{numb}0}}}{1}
      {1}{{{\color{numb}1}}}{1}
      {2}{{{\color{numb}2}}}{1}
      {3}{{{\color{numb}3}}}{1}
      {4}{{{\color{numb}4}}}{1}
      {5}{{{\color{numb}5}}}{1}
      {6}{{{\color{numb}6}}}{1}
      {7}{{{\color{numb}7}}}{1}
      {8}{{{\color{numb}8}}}{1}
      {9}{{{\color{numb}9}}}{1}
      {:}{{{\color{punct}{:}}}}{1}
      {,}{{{\color{punct}{,}}}}{1}
      {\{}{{{\color{delim}{\{}}}}{1}
      {\}}{{{\color{delim}{\}}}}}{1}
      {[}{{{\color{delim}{[}}}}{1}
      {]}{{{\color{delim}{]}}}}{1},
}

\usepackage{fbox}
\newcommand{\rqsummary}[2]{
        \vspace{2mm}
        \noindent
        \fbox{%
            \parbox{.97\linewidth}{%
                    \textbf{#1 Summary.}
                #2
            }%
        }%
        \vspace{2mm}
}%

\newcommand{\rqinterpretation}[2]{
        \vspace{0.4em}
        \noindent
        \colorbox{gray!40}{%
            \parbox{.97\linewidth}{%
                    \textbf{#1 Lesson Learned.}
                #2
            }%
        }%
}%

\newcommand{\passau}{University of Passau\xspace}
\newcommand{\reutlingen}{Reutlingen University (Herman Hollerith Zentrum, HHZ)\xspace}

\usepackage[colorinlistoftodos,prependcaption,textsize=tiny]{todonotes}

\definecolor{Gray}{gray}{0.9}

\newcommand\zz[1]{%
\ifdim#1pt<1pt\cellcolor{white}\else
\ifdim#1pt<5pt\cellcolor{gray!10}\else
\ifdim#1pt<10pt\cellcolor{gray!20}\else
\ifdim#1pt<15pt\cellcolor{gray!30}\else
\ifdim#1pt<20pt\cellcolor{gray!40}\else
\ifdim#1pt<25pt\cellcolor{gray!50}\else
\ifdim#1pt<30pt\cellcolor{gray!60}\else
\ifdim#1pt<40pt\cellcolor{gray!70}\else
\ifdim#1pt<50pt\cellcolor{gray}\else
\ifdim#1pt>50pt\cellcolor{white}\else
\cellcolor{white}\fi\fi\fi\fi\fi\fi\fi\fi\fi\fi
#1}
\newcolumntype{C}{>{\collectcell\zz}c<{\endcollectcell}}

\newcommand\rr[1]{%
\ifdim#1pt<1pt\cellcolor{white}\else
\ifdim#1pt<2pt\cellcolor{gray!10}\else
\ifdim#1pt<3pt\cellcolor{gray!20}\else
\ifdim#1pt<4pt\cellcolor{gray!30}\else
\ifdim#1pt<5pt\cellcolor{gray!40}\else
\ifdim#1pt<6pt\cellcolor{gray!50}\else
\ifdim#1pt<7pt\cellcolor{gray!60}\else
\ifdim#1pt<8pt\cellcolor{gray!70}\else
\ifdim#1pt<9pt\cellcolor{gray}\else
\cellcolor{white}\fi\fi\fi\fi\fi\fi\fi\fi\fi
#1}
\newcolumntype{R}{>{\collectcell\rr}c<{\endcollectcell}}

\usepackage{amsmath}
\usepackage{cite}

\renewcommand*\descriptionlabel[1]{%
\hspace\labelsep\normalfont #1}

\title{Exposing Software Engineering Students to Stressful Projects: Does Diversity Matter?}

\begin{document}

\author{%
	\IEEEauthorblockN{Isabella Gra{\ss}l}%
	\IEEEauthorblockA{University of Passau\\
	Passau, Germany\\
	isabella.grassl@uni-passau.de}
\and
	\IEEEauthorblockN{Gordon Fraser}%
	\IEEEauthorblockA{University of Passau\\
	Passau, Germany\\
	gordon.fraser@uni-passau.de}
\and
	\IEEEauthorblockN{Stefan Trieflinger}%
	\IEEEauthorblockA{Reutlingen University\\
	Reutlingen, Germany\\
	stefan.trieflinger@reutlingen-university.de}
\and
	\IEEEauthorblockN{Marco Kuhrmann}%
	\IEEEauthorblockA{Reutlingen University\\
	Reutlingen, Germany\\
	kuhrmann@acm.org}
}

\maketitle

\begin{abstract}
  Software development teams have to face stress caused by deadlines,
  staff turnover, or individual differences in commitment, expertise,
  and time zones. While students are typically taught the theory of
  software project management, their exposure to such stress factors
  is usually limited. However, preparing students for the stress they
  will have to endure once they work in project teams is important for
  their own sake, as well as for the sake of team performance in the
  face of stress.
  Team performance has been linked to the diversity of software
  development teams, but little is known about how diversity
  influences the stress experienced in teams.
  In order to shed light on this aspect, we provided students with the
  opportunity to self-experience the basics of project management in
  self-organizing teams, and studied the impact of six diversity dimensions on team
  performance, coping with stressors, and positive perceived learning effects.
  Three controlled experiments at two universities with a total of 65
  participants suggest that the social background impacts the
  perceived stressors the most, while age and work experience have the highest impact on perceived learnings. Most diversity dimensions have a medium correlation with the quality of work, yet no significant relation to the team performance. 
  This lays the foundation to improve students' training for software engineering teamwork based on their diversity-related needs and to create diversity-sensitive awareness among educators, employers and researchers.
\end{abstract}



\section{Introduction}
\label{sec:intro}

Modern software development is a collaborative task conducted by self-organizing cross-functional teams \cite{hilderbrand2020b,storey2020b}. This aspiration, which is also reflected in the ``Agile Manifesto'' \cite{beck2001manifesto}, has become an ideal pursued by many companies and, therefore, educating students to develop software collaboratively in small projects has become commodity \cite{chatley2017,hussain2020,butt2022}. Besides the technical dimension, i.e., programming techniques and tools, the human component has to be taken into account as well~\cite{perry1994,storey2020b}. Yet, during the Covid-19 pandemic, face-to-face team work has been replaced with online work, home office, and distance learning, which also changed the ways students were educated in the past two years \cite{means2020,wildman2021}. After returning to face-to-face labs, students now have to deal with other people again. Students need to learn again how to ``function" in a team, and how to deal with group dynamics and stressful situations~\cite{wildman2021,stevenson2006,curseu2013a,gustems-carnicer2019}, which is hard to implement in classroom settings. Project work is time-consuming, actual development tasks are likely to draw away students' attention from the management activities \cite{kuhrmann2016a,mas2021}, and students have limited opportunities to gain experience in the key concepts of, e.g., collaboration, communication, coordination, and documentation \cite{haller2000,paasivaara2018}. 

At the same time, project work as such has changed: It has become global, leading to new challenges related to teams \cite{10.1007/978-3-540-73460-4_6,10.1145/1852786.1852817,DBLP:journals/software/EbertKP16}. Co-located as well as distributed teams are diverse (e.g., different cultures or educational backgrounds), and several studies indicated that diverse teams are more productive~\cite{catolino2019c,russo2020e} and lead to socio-cultural advantages, e.g., improved workplace atmosphere and decreased delinquency \cite{blincoe2019c,russo2020e,wang2020d,gardenswartz2003a}. Students therefore need to be educated with respect to the ever-increasing demand for diversity in software engineering (SE) at all levels---education, academia, and industry~\cite{albusays2021,greifenstein2021}. Yet, it is a challenge to provide students with an environment in which they are exposed to stressful situations, allowing them to learn coping with challenging situations in diverse teams~\cite{gefen2012}. 

\paragraph*{Objective}
Our overall objective is to study group dynamics in SE education focussing on two aspects: (i) working as a team under stress and (ii) working in diverse teams. We aim to study whether students of different diversity dimensions---age, sex, ethnicity, education, social background, and work experience---experience stressors and learnings of teamwork under stress differently, whether diverse teams experience those factors differently, and whether they perform differently than more homogeneous teams.

\paragraph*{Contribution}
Using a controlled experiment in which students work in diverse teams that are exposed to different stressors while working on a straightforward task focussing on social skills rather than technical skills, we analyze the teams' performance as well as the students' experiences regarding the stressors and personal learnings in relation to six diversity dimensions. To provide a transparent and reproducible framework for the experiment, we used an available experiment design~\cite{DBLP:conf/icse/KuMue16} for measuring student behavior in stressful team scenarios, for which a replication package was available, and implemented the experiment thrice in two undergraduate SE courses at \reutlingen and the \passau (UP).
Our results indicate that the diversity dimension of \textit{social background} affects stressors the most, while \textit{age} and \textit{work experience} have the greatest correlation with the perceived learnings. 
Teams with a high variety of age, ethnicities and work experiences seem to perceive fewer stressors, yet more learnings. Considering the quality of work, we find a medium correlation for most diversity dimensions, although we could not find a significant relation between the dimensions of diversity and team performance.
These results provide educators and researchers with an improved understanding of the experiences of students and, thus, how to better prepare them for their professional life as a software developer while creating diversity-sensitive awareness. 

\paragraph*{Outline}
The remainder of the paper is organized as follows:~\cref{sec:relatedwork} provides an overview of related work. In~\cref{sec:methodology}, we describe the research design, before we present the results and interpretations of our study in~\cref{sec:results}, and conclude the paper in~\cref{conclusions}.


\section{Related Work}
\label{sec:relatedwork}
In this section, we address stress in general, in professional teamwork, and in student teams---all in the context of diversity.

\subsection{Stress and Diversity in General}
Diversity means distinguishing characteristics of people at several layers \cite{gardenswartz2003a}: \emph{internal} diversity (e.g., age), \emph{external} diversity (e.g., work experience), and \emph{organizational} dimensions (e.g., work location). In our research, we focus on six selected diversity dimensions, which are relevant for SE \cite{rodriguez-perez2021b} and that are ethically justifiable.
Research shows that certain groups of people respond to stressful situations with similar behavioral patterns differing from other groups of people \cite{arthur1998,brougham2009,kock2018,ralph2020,russo2020b}.
For instance, Gustems-Carnicer et~al.~\cite{gustems-carnicer2019} studied the dimensions \emph{age} and \emph{working setting} and found that older student teachers are less affected by stress and performance. They suggest that student teachers should already be exposed to stressful situations at university to improve their training and coping strategies. 
Several studies found that males and females have different coping strategies for different stressors~\cite{arthur1998, brougham2009, gefen2012}. Furthermore, female students' stress level increases more rapidly and levels off, while for males the stress level increases more slowly but is higher over time than for females---both at university and in the private space. 

The study described in this paper addresses two gaps in literature: (1) to the best of our knowledge, stress and diversity of SE students have not been explicitly studied, and (2) most generic stress studies focus on \emph{age}, \emph{gender}, or \emph{ethnicity}, but rarely address \emph{external} diversity, e.g., parental status or educational background \cite{gardenswartz2003a}.

\subsection{Stress and Diversity in Professional SE Teams}
Agility and globally distributed software development fostered research on heterogeneous software teams \cite{7577436,7976681,storey2020b,albusays2021}. Such teams are per se diverse, and it is assumed that the different experience, skills, and mindsets serves a greater variety of knowledge and perspectives, and thus diverse teams plan more effectively and find more creative solutions \cite{harrison2007s, van2007}. However, diversity can also lead to higher potential for conflict or lower group cohesion \cite{van2007}.
Nonetheless, collaboratively developing a software solution under pressure, e.g., time-to-market or meeting feature/quality requirements, is key to succeed in competitive scenarios \cite{hilderbrand2020b, ciancarini2019}. This puts teams under stress, which is a multifaceted factor. For example, Tehrani~\cite{tehrani2005managing} identified the following stressors: \emph{``unsympathetic organizational culture, poor communication between managers and employees, lack of involvement in decision-making, bullying and harassment, continual or sudden change, insufficient resources, conflicting priorities and lack of challenge''}, which are often perceived very negatively and can lead to frustration, which in turn especially affects underrepresented groups as they often feel intimidated and not heard \cite{filippova2017a}.
Even though the determining factor is often human, large-scale field studies on the effects that diversity can have on team productivity and well-being are currently still missing \cite{ralph2020,russo2020b}. Furthermore, most studies only consider the dimensions \emph{sex/gender} or \emph{ethnicity}. For example, women may have positive effects in teams and mitigate certain risk factors \cite{russo2020e}. Similarly, the presence of women in teams may reduce community smells \cite{catolino2019c}, and most studies suggest that heterogeneous teams are more effective than homogeneous ones because of improved communication \cite{ciancarini2019}.
There are, however, many more aspects of diversity \cite{gardenswartz2003a}.

In this paper, we address a gap in the existing literature by providing a more holistic picture: We study stress, learnings, and performance of student teams in the light of multiple dimensions of diversity.

\subsection{Stress and Diversity in SE Student Teams}

Since SE deals with socio-technical systems developed by mostly interdisciplinary teams, it is key to teach students social and organizational skills \cite{perry1994, storey2020b, herro2017}. For this, project-based learning provides positive learning effects \cite{shakerihosseinabad2019, peraire2019, huang2008, bavota2012, ralph2018, chatley2017, butt2022,khmelevsky2016,ikonen2009}, specifically regarding communication and collaboration, which are important in stressful environments \cite{mas2021,paasivaara2018}. In such situations, older students and experienced students seem to react more calmly to stress~\cite{kock2018}. Cen et~al.~\cite{cen2014a} found that all-female groups outperformed all-male groups and, at the individual member level, mixed-group students improved their engagement and collaborative learning the most. Hence, the experience of teamwork is often positively influenced by having female members in teams~\cite{corbacho2021a}. Communication and collaboration in diverse teams in stressful situations constitute big challenges and, therefore, requires accurate training, especially for teams with different backgrounds \cite{stevenson2006, curseu2013a}.
This challenge became even more evident during the Covid-19 pandemic that forced a shift to online teaching, challenged teamwork \cite{ralph2020}, and introduced new stressors to students \cite{means2020} related to communication as well as task and role allocation within student projects \cite{wildman2021}. 

In this study, we analyze for the first time the stressors, learnings, and performance with respect to six dimensions of diversity in SE student teams to address a gap in the literature regarding diversity-dependent effects of stress in teams.



\section{Research Method}
\label{sec:methodology}


In the following, we present our research questions~(\Cref{sec:Method:ObjectiveAndRQs}), the experiment design~(\Cref{sec:Method:ExperimentDesign}), the data collection (\Cref{sec:Method:DataCollection}), and the data analysis (\Cref{sec:Method:DataAnalysis}).

\subsection{Research Objective and Research Questions}
\label{sec:Method:ObjectiveAndRQs}
Our overall objective is to study individual experiences and team performance of student teams in stressful situations considering different dimensions of diversity. Specifically, we aim to answer the following research questions:


\smallskip
\noindent\textbf{RQ1:} \emph{What do students consider stressors in team work and are those stressors affected by different diversity dimensions?} There are many stressors, notably, for novices under time pressure in an unfamiliar team working on an underspecified task. This research question aims at identifying the stressors students experience as individuals while working as a team.

\smallskip
\noindent\textbf{RQ2:} \emph{What do students consider learnings in team work and are those learnings affected by different diversity dimensions?} Challenges in stressful situations might lead to behavior that implies positive learning. In this research question, we aim at identifying such learnings of the individuals.

\smallskip
\noindent\textbf{RQ3:} \emph{Does diversity affect the perceived stressors and learnings in teams?} Having identified stressors and learning at the individual level, in this research question, we aim at studying stressors and learnings at the team level, i.e., whether teams experience stressors and learnings differently depending on the various dimensions of diversity.

\smallskip
\noindent\textbf{RQ4:} \emph{Does diversity affect the performance under stress?}  While RQ1--RQ3 focus on personal experiences, we also aim to study whether differences in the performance and the quality of work exist and how heterogeneous teams perform in comparison to homogenous teams.


\subsection{General Experiment Design}
\label{sec:Method:ExperimentDesign}
\begin{figure}[!t]
	\centering
	\includegraphics[width=\linewidth]{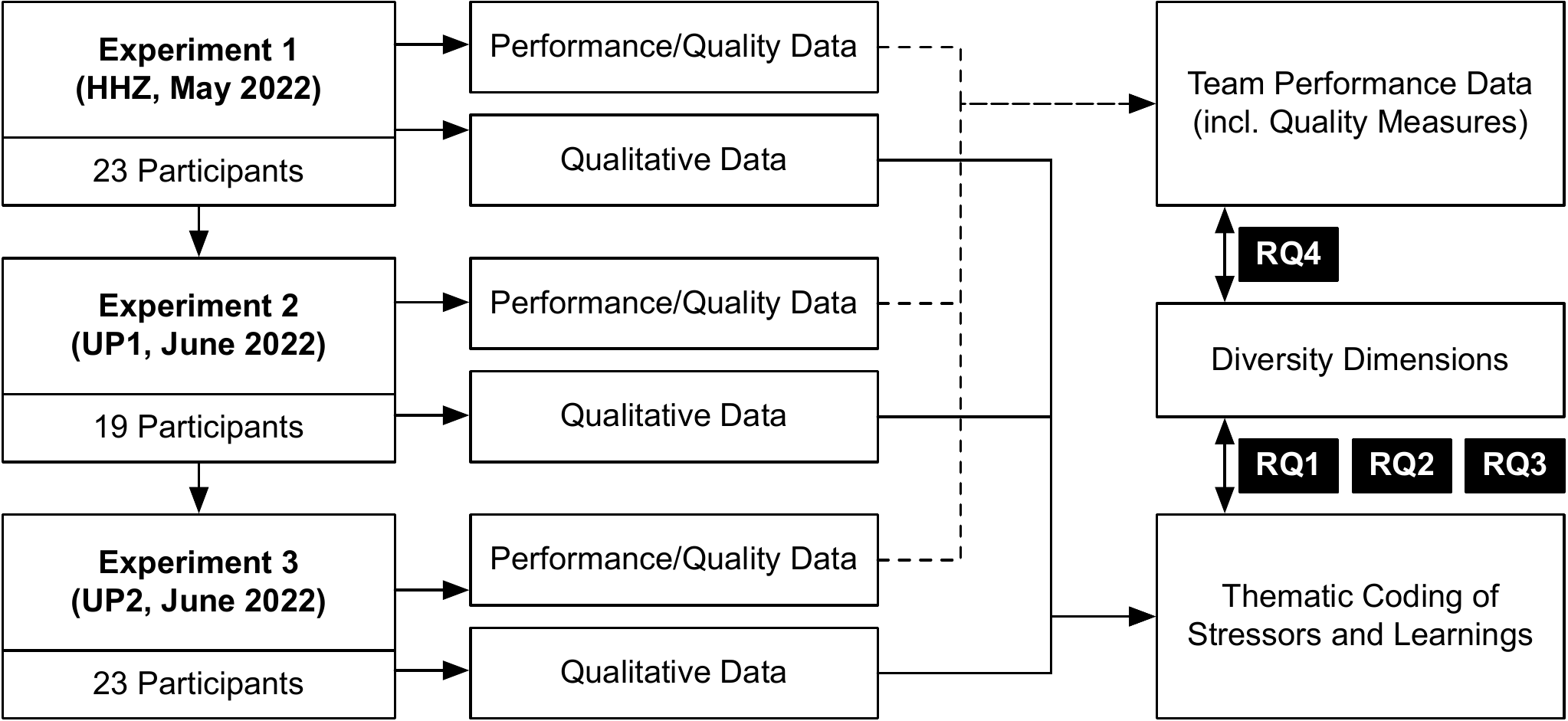}
	\caption{Overview of the research method of this study.}
	\label{fig:OverallResearchDesign}
\end{figure}

The overall research was organized according to the schema shown in~\Cref{fig:OverallResearchDesign}. To analyze the team performance, quality of work, stressors and learnings in relation to the different diversity dimensions, we adopted the experiment design as proposed and described in detail in Kuhrmann and M\"unch~\cite{DBLP:conf/icse/KuMue16},
which defines a very simple task---sorting and counting sweets. 
We opted for this design as it allows for focussing on our study aim, the teams and the group dynamics, by taking out all actual software implementation tasks, thus avoiding students' attention being drawn away. 

\subsubsection{Overview}
\label{sec:Method:ExperimentDesign:Overview}
\begin{figure*}[t]
	\centering
	\includegraphics[width=\linewidth]{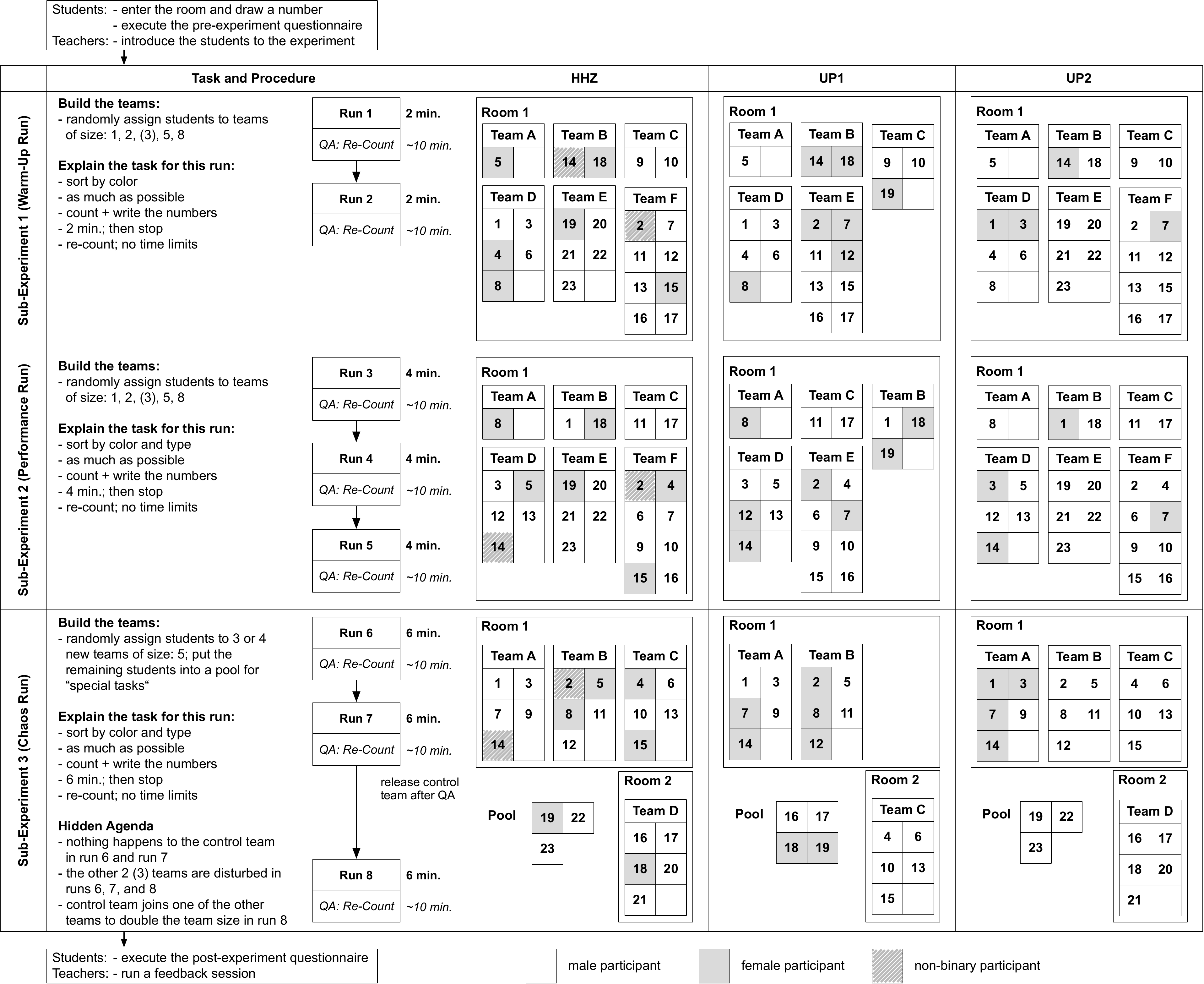}
	\caption{Overview of the experiment as implemented at the different sites (incl.\ general organization, task overview, subject, participant IDs and room assignments). Every run generates two controlling sheets per team to measure the team performance.}
	\label{fig:experiment}
\end{figure*}

The task was implemented using the \emph{controlled experiment} instrument as described by Wohlin et~al.~\cite{wohlin2012experimentation}. \Cref{fig:experiment} provides an overview of the whole experiment, which was implemented thrice at two universities. The experiment consists of three sub-experiments in eight runs, and takes approx.\ 3 hours in total to complete. The three sub-experiments serve different purposes: sub-experiment~1 is the \emph{warm-up run} in which the students familiarize themselves with the general task. Sub-experiment~2 is the \emph{performance run} in which the students have to fulfill a slightly more complex task and have three runs instead of two. Finally, sub-experiment~3 is the \emph{chaos run} in which the students execute the same task as in sub-experiment~2, but are exposed to ``special tasks'' such as several interventions, e.g., \emph{``Go outside''} or \emph{``Eat some counted/sorted items''}, and stressors to which the students have been exposed, e.g., time pressure, resource bottlenecks, and underspecified work approach definitions. 
For each sub-experiment, students were randomly assigned to teams.

\subsubsection{Participants}
\label{sec:Method:ExperimentDesign:Procedure}
We implemented this experiment in three sessions embedded in two undergraduate SE courses. The first experiment was implemented in the course ``Empirical Methods for Software Engineering'' at \reutlingen in May 2022, as a course-integrated project activity for the students of this course. The other two experiments took place in the course ``Introduction to Software Engineering'' at the \passau (UP) in June 2022. The participation was voluntary, but participating students were given a bonus for the final exam. After each experiment, all students were informed about the purpose of the study and gave their consent about the publication process.
\begin{table}[t]
\renewcommand{\arraystretch}{0.85}
\footnotesize
\caption{Diversity dimensions from self-reported demographics of the students from all experiments (n=65).}
\label{tab:StudentDemographics}
	\begin{tabularx}{\linewidth}{p{0.16\linewidth}Xrrrr}
	\toprule
	\multicolumn{2}{l}{Category} & HHZ & UP1 & UP2 & $\sum$ \\ 
	\midrule
	Age & $<$21    & 14 &  7 & 14 & 35 \\
	    & $\geq$21 &  9 & 12 &  9 & 30 \\
	\midrule
	Sex & female        &  6 &  7 &  4 & 17 \\
	    & male          & 15 & 12 & 19 & 46 \\
	\midrule
	Ethnicity & local & 20 & 18 & 18 & 56  \\
	          & non-local &  3 &  1 &  5 &  9  \\
	\midrule
	Education & high school & 21 & 17 & 21 & 59 \\
	          & other      &  2 &  2 &  2 &  6 \\
	\midrule
	\multirow[t]{3}{=}{Social background} & both parents aca. &  5 & 10 & 7 & 22 \\
	       & one parent aca.    &  2 &  4 & 8 & 14 \\
	       & no parent aca.     & 16 &  5 & 8 & 29 \\
	\midrule 
	\multirow[t]{2}{=}{Work experience} 
		& $<$6 months        & 11 & 8  & 16 & 35 \\
	     	& $\geq$6 months  & 12 & 11 & 7  & 30 \\
	\bottomrule
\end{tabularx}
\end{table}

\Cref{tab:StudentDemographics} provides an overview of the 65 students of which 26.2\% reported their biological sex as female and 70.8\% as male. A majority of the participants (86.1\%) originated from Germany where both universities are located and which we refer to as \emph{local}, while \emph{non-local} refers to students from other countries. The average age was 21 years, and the majority of the students (90.7\%) entered the academic education with a high school degree. Regarding their social background, it is noticeable that the majority of the HHZ students came from households in which the parents have no academic education, which differs from the UP students who, to a large extent, came from households in which at least one parent has an academic education. To assess the students' knowledge, we asked them to report their work experience, including work as a student assistant. In total, 53.8\% of the students had less than six months of work experience.

\subsubsection{Procedure}
\label{sec:Method:ExperimentDesign:Procedure}

At the very beginning, students entered the room and drew a number, which served as identifier in the experiment. Having drawn their numbers, students completed the pre-experiment questionnaire (diversity dimensions; \Cref{tab:StudentDemographics}). The educators introduced the students to the experiment and presented the first group assignment. As the room was set up accordingly~(\Cref{fig:experiment}), i.e., tables each with team identifier, chairs, and one pen, students got together in their teams while the educators provided the initial controlling sheets and the items to sort. Eventually, the task was introduced to the students as follows: \emph{``The following items need to be sorted by color (and type). You can decide yourselves in which way you work. The only thing that matters is that your team sorts as many items as possible. Document your outcomes using the provided controlling sheets.''} 
When the preparation was completed, educators started the first run of the first sub-experiment. Subsequently, the experiment was implemented following the procedure illustrated in \Cref{fig:experiment}. After the experiment's completion, all students were summoned in one room, filled out the post-experiment questionnaires, and participated in the debriefing that, among other things, included a retrospective, a feedback and Q\&A session, and an explanation of the ``hidden agenda''.


\subsection{Data Collection Procedures}
\label{sec:Method:DataCollection}
We collected quantitative and qualitative data. To collect quantitative data, the following procedure was implemented: Before a run started, teams were provided with the first controlling sheet that had to be used to document the numbers during the run which contains an empty table with the columns color, type and number of the counted sweets. When the run's time elapsed, the sheets were \emph{immediately} collected and teams were provided with a fresh sheet to re-count the sorted items without pressure. That is, for each run $i$, every team $t$ filled out two controlling sheets to collect the performance data $\text{wi}_{t,i}$ under and $\text{ci}_{t,i}$ without time pressure. 

To collect qualitative data, students were provided with a post-experiment questionnaire after the final run, which consisted of two parts with open-ended free-text questions: the first part asked for things that went well in the three sub-experiments, and part two asked for things that did not (up to six statements in total). The students were asked to anonymously reflect and document their experience and to provide some rationale.  
%

\subsection{Data Analysis Procedures}
\label{sec:Method:DataAnalysis}

Quantitative data was transcribed into a spreadsheet for further analyses using R. One researcher transcribed the data, a second researcher did a quality check, and a third researcher performed that actual analyses. Qualitative data from the post-experiment questionnaire was also transcribed into spreadsheets that served for the qualitative analysis\footnote{For replications, all data of the study are available at https://figshare.com/s/0cc2384af92bc3dd198e. All sensitive demographic data are available upon request.}. 

\paragraph*{Analysis Procedure for RQ1 and RQ2}
To answer RQ1 and RQ2, we analyzed the coded students' responses regarding positive and negative aspects of the experiment extracted from the post-ex\-pe\-ri\-ment questionnaires to which we applied a thematic analysis \cite{bergman2010hermeneutic}:  For each question, (i) we collected themes emerged from the codes, (ii) counted the themes, and (iii) related the themes to the data and our research questions. To ensure inter-rater reliability, two researchers independently classified the statements for each open question and, eventually, agreed on a coding scheme. Each rater then classified all statements with an inter-rater agreement of Cohen's $\kappa = 0.87$, which is considered very reliable~\cite{dewinter2016}.

To analyze the coded data, we calculated the percentage of the codes for each of the six diversity dimensions. We used the \emph{Mann-Whitney U}-test at a significance level of $\alpha\leq0.05$ to measure statistical differences within the different diversity dimensions. Since only two students identified themselves as neither female nor male, we excluded their codes from the dimension \emph{sex}. 
For the non-binary category \emph{social background}, we adjusted the significance level with a \emph{Bonferroni correction}~\cite{Miller:2013aa} by dividing the given $\alpha$ by the number of tests, which resulted in a corrected $\alpha_{\text{B\_cor}}\leq0.016$. 
To measure the rank correlation between both stressors (RQ1) and perceived learnings (RQ2), we used \emph{Spearman's} rank correlation at a significance level of $\alpha\leq0.05$. 
\paragraph*{Analysis Procedure for RQ3}
To answer RQ3, we analyzed the stressors and perceived learnings obtained in RQ1 and RQ2 in relation to the six diversity dimensions. For this, we pose the hypotheses: $\text{H1}_{0}$: \emph{There is no relation between identified stressors and a team's diversity} and $\text{H2}_{0}$: \emph{There is no relation between stated learnings and a team's diversity}.

%

As this analysis focusses on teams instead of individuals, we computed the diversity scores of the teams using the Blau index~\cite{blau1977}. The index was computed for each diversity dimension for the 11 teams in the third sub-experiment (run 3.6 and 3.7). These teams are comparable since at this point all participants know the task and the conditions of the experiment, and have experienced working in random teams. 

Having computed the Blau index, we filtered the qualitative data, such that those stressors and learnings were extracted that have been mentioned by the 11 selected teams for the two experiment runs of interest. We tested for correlations between the six diversity dimensions and the stressors, and we tested for correlations between the six diversity dimensions and the perceived learnings using Pearson's \emph{r}. 
As we executed six tests each for the stressors and the learnings, we used \emph{Bonferroni correction} again to adjust the significance level, which resulted in a corrected $\alpha_{\text{B\_cor}}\leq0.008$.

\paragraph*{Analysis Procedures for RQ4}
To answer RQ4, we used the data from the controlling sheets together with the Blau index computed for RQ3, and pose hypothesis $\text{H3}_{0}$: \emph{There is no relation between team performance and a team's diversity}. 

Based on the performance data $\text{wi}_{t,i}$ and $\text{ci}_{t,i}$, we computed the relative error $\text{err}_{\emph{rel}}$ as share of errors for counted items and the clerical error $\text{err}_{\emph{cl}}$\footnote{A clerical error occurs if students wrote down ``wrong'' numbers. Clerical errors are used as share of errors made in total. For instance, if in a run 20 item types are listed and for 10 item types wrong values have been reported, the total clerical error is 10, i.e., 50\% of all errors made were clerical errors.}. 
A team's performance for the two runs of interest is then characterized by the mean of re-counted items (counting performance) together with the mean of the clerical error (quality of work). To answer RQ4, we used Pearson's \emph{r} following the exact same procedure as applied to RQ3. However, since we used 12 combined tests, we used a \emph{Bonferroni}-corrected significance level of $\alpha_{\text{B\_cor}}\leq0.004$.

\subsection{Threats to Validity}
\label{sec:Discussion:TTV}


We discuss potential threats to validity using the categories by Wohlin et~al.~\cite{wohlin2012experimentation}. 
The design of the study including the questionnaire is approved by a member of the ethics committee of the \passau. The construct focuses on studying selected aspects of group dynamics. For this, the original experiment instrument \cite{DBLP:conf/icse/KuMue16} was thoroughly reviewed for applicability by three researchers. Modifications of the original design only concern the runs' duration in the third sub-experiment (6 minutes instead of 8 minutes) to increase the stress level. The data analysis procedures were conducted and reviewed by three independent researchers to ensure internal validity. Students may influence each other's behavior and responses, as well as instructor performance. Therefore, several additional female and male experts of SE and didactics ensured the smooth and identical performance of the experiments. 
Regarding the external validity, this study has been conducted several times at several universities. Also, the general experiment setup follows the original setup with only one modification (see above). Even though we also found all originally mentioned stressors in our experiment, it needs to be mentioned that the experiment design is explicitly targeted to group dynamics. Thus, one limitation of the experiment is that other stress-related factors, e.g., complaining customers or re-work of software artifacts, are not part of the experiment. 
However, due to the focused design the experiment is universal and we provide all the materials to enable replications of the study and to incrementally grow a knowledge base.



\section{Results}
\label{sec:results}
In this section, we present the results and their interpretation structured according to the research questions.

\subsection{RQ1: Stressors and Diversity}
\label{sec:Results:RQ1}
\subsubsection{Extraction and Analysis of Stressors}
Based on the post-experiment questionnaire, we analyzed 298 mentions and extracted 19 stressors.
%
\Cref{tab:heatmap-stressors-runs} shows the absolute numbers of mentioned stressors for each sub-experiment across all three experiments. 
To provide a structured analysis, we defined the three top-level categories \emph{organizational factors} (75 mentions), \emph{planning \& strategy} (160), \emph{affective factors} (55), and \emph{other} (8).

\begin{table}[tb]
	\centering
	\vspace{-0.5em}
	\caption{Students' mentioned stressors for each sub-exp.}
	\vspace{-0.5em}
	\label{tab:heatmap-stressors-runs}
    	\begin{tabularx}{\linewidth}{lXCCCr}
	\toprule
    	Category & Subcategory &  \multicolumn{1}{c}{HHZ} & \multicolumn{1}{c}{UP1} & \multicolumn{1}{c}{UP2} & \multicolumn{1}{c}{$\sum$} \\ 
    	\midrule
        Organisat.  & Team Organization & 3 & 6 & 0 & 9 \\ 
        Factors & Bigger Teams' Ineffec. & 5 & 4 & 6 & 15 \\ 
        ~ & Smaller Teams' Ineffec. & 5 & 6 & 1 & 12 \\ 
        ~ & Loss of Key Pers. &  0 & 0 & 3 & 3 \\ 
        ~ & Staff Turnover &  0 & 1 & 7 & 8 \\ 
        ~ & Ext. Disturbances &  3 & 0 & 25 & 28 \\ 
        Planning \&  & Missing Work Appr. & 31 & 5 & 5 & 41 \\ 
        Strategy & Ineffective Appr. & 8 & 9 & 6 & 23 \\ 
        ~ & No Common Appr. & 7 & 5 & 4 & 16 \\     
        ~ & Task Distribution & 2 & 2 & 2 & 6 \\ 
        ~ & Time Management & 18 & 9 & 3 & 30 \\ 
	 & Time & 21 & 10 & 1 & 32 \\         
        ~ & Overloading &  2 & 3 & 0 & 5 \\ 
        ~ & Bottleneck (Res.) &  2 & 2 & 3 & 7 \\          
        Affective  & Boring Task & 1 & 1 & 7 & 9 \\ 
        Factors & Lack of Motivation & 1 & 1 & 8 & 10 \\ 
        ~ & Communication & 2 & 6 & 6 & 14 \\ 
	~ & Task Complex. & 0 & 18 & 2 & 20 \\ 
        ~ & Noise Level &  1 & 1 & 0 & 2 \\ 	        
        Other & ~ & 0 & 1 & 7 & 8 \\         
        $\sum$ & ~ & 112 & 90 & 96 & 298 \\ 
        \bottomrule
    \end{tabularx}
\end{table}

The analysis shows that students detect the majority of the stressors at different stages of the experiment. In the first experiment, the most frequently mentioned stressors were \emph{missing work approach} (31), \emph{time} (21), and \emph{time management}~(18). Over time, the perception of stressors changed. In the second sub-experiment, the \emph{time} (10) issue started to fade out, whereas the \emph{task complexity} (18) posed a new challenge. Eventually, most stressors mentioned in the first two sub-experiments were not mentioned that frequently in the third sub-experiment, i.e., students learned to deal with the inherent stress factors, yet, were challenged by those factors actively applied using the explicit treatments (\emph{external disturbances}: 25). Noticeable, the factor \emph{noise level} does not seem to be considered a stressor by the students, even though it actually was fairly noisy, especially in the third sub-experiment.

\subsubsection{Relationships between Stressors}
\begin{figure}[t]
	\centering
	\includegraphics[width=\linewidth]{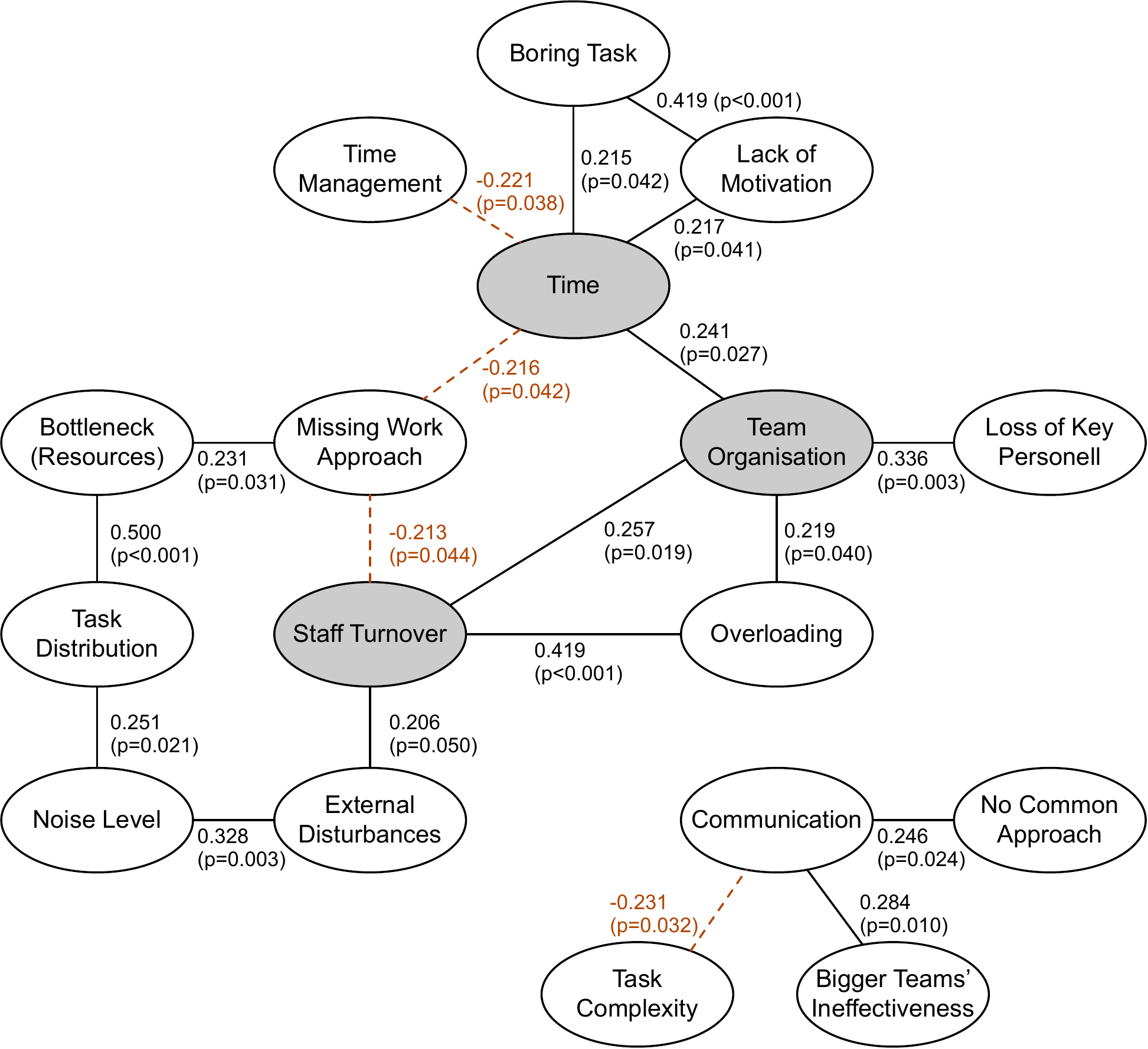}
        \vspace{0.5em}
	\caption{Linear relation between the stressors.}
	\label{fig:Results:RQ1:LinearStressors}
\end{figure}

Since the students were exposed to several stressors, we analyzed the mentioned stressors for relationships. \Cref{fig:Results:RQ1:LinearStressors} visualizes those linear relations between the stressors that are significant. The analysis shows that (i) 17 out of the 19 analyzed stressors are present in the dependency graph, and (ii) \emph{staff turnover}, \emph{time}, and \emph{team organization} form centers with multiple significant relations, i.e., they can be considered key stress factors in the experiment.
Especially the negative relations between \emph{time}, \emph{time management}, \emph{staff turnover}, and \emph{missing work approach} show that time pressure and the need to develop a work approach in an ``unstable team'' constitute considerable challenges for the participants. The stressors \emph{ineffective approach} and \emph{smaller teams' ineffectiveness} were mentioned by the participants, although there are no significant linear relations to any of the other stressors.

\begin{table*}[tb]
	\centering
	\vspace{-0.5em}
	\caption{Percentage of mentioned stressors from the students belonging to a certain dimension of diversity.}
	\vspace{-0.5em}
	\label{tab:heatmap-stressors-diversity}
	\begin{tabularx}{\linewidth}{XlCCCCCCCCCCCCC}
    	\toprule
        Category & Subcategory & \multicolumn{2}{c}{Age}  & \multicolumn{2}{c}{Sex}  & \multicolumn{2}{c}{Ethnicity}  & \multicolumn{2}{c}{Education}& \multicolumn{3}{c}{Social BG} & \multicolumn{2}{c}{Work}  \\ 
        ~ & ~ & \multicolumn{1}{c}{$<$21} & \multicolumn{1}{c}{$\geq$21} & \multicolumn{1}{c}{f} & \multicolumn{1}{c}{m} & \multicolumn{1}{c}{loc.} & \multicolumn{1}{c}{non-l.}& \multicolumn{1}{c}{high.} & \multicolumn{1}{c}{oth.} & \multicolumn{1}{c}{both} & \multicolumn{1}{c}{one} & \multicolumn{1}{c}{none} & \multicolumn{1}{c}{$\geq$0.5y} & \multicolumn{1}{c}{$<$0.5y} \\ 
        \midrule
        Organizat.& Team Orga. & 4.76 & 4.44 & 5.88 & 4.35 &  4.17 & 7.41 & 4.52 & 5.56 & 1.52 & 7.14 & 5.75 & 3.33 & 5.71 \\ 
        Factors & Bigger Teams' Ineff. & 10.48 & 4.44 & 5.88 & 7.97 &  7.14 & 11.11 & 7.34 & 11.11 & 6.06 & 9.52 & 8.05 & 6.67 & 8.57\\ 
        ~ & Smaller Team' Ineff. & 6.67 & 5.56 & 3.92 & 7.25 & 5.95 & 7.41 & 6.78 & 0.00 & 3.03 & 9.52 & 6.90 & 5.56 & 6.67  \\         
	~ & Loss of Key Person. & 0.95 & 2.22 & 1.96 & 1.45 &  1.79 & 0.00 & 1.69 & 0.00 & 1.52 & 2.38 & 1.15 & 2.22 & 0.95  \\ 
         ~ & Staff Turnover & 4.76 & 3.33 & 3.92 & 4.35 &  4.17 & 3.70 & 4.52 & 0.00 & 0.00 & 4.76 & 6.90 & 4.44 & 3.81  \\ 
        ~ & Ext. Disturbances &13.33 & 15.56 & 13.73 & 15.22 & 15.48 & 7.41 & 14.69 & 11.11 & 16.67 & 7.14 & 16.09 & 18.89 & 10.48 \\ 
  
        Planning, & Missing Work Appr. & 19.05 & 23.33 & 19.61 & 20.29 &  22.62 & 11.11 & 20.90 & 22.22 & 28.79 & 19.05 & 16.09 & 23.33 & 18.05  \\ 
	Strategy & Ineffective Appr. & 11.43 & 12.22 & 17.65 & 10.14 & 10.71 & 18.52 & 12.43 & 5.56 & 9.09 & 11.90 & 13.79 & 7.78 & 15.24  \\ 
        ~ & No Common Appr. & 8.57 & 7.78 & 9.80 & 7.97 & 8.93 & 3.70 & 7.91 & 11.11 & 13.64 & 7.14 & 4.60 & 4.44 & 11.43  \\ 
        ~ & Task Distribution & 0.95 & 5.56 & 3.92 & 2.90 & 3.57 & 0.00 & 2.82 & 5.56 & 4.55 & 7.14 & 0.00 & 4.44 & 1.90  \\ 
        ~ & Time Management & 18.10 & 12.22 & 23.53  & 12.32 &16.67 & 7.41 & 15.82 & 11.11 & 16.67 & 14.29 & 14.94 & 16.67 & 14.29 \\ 
	~ & Time & 12.38 & 21.11 & 23.53 & 13.77  & 14.29 & 29.63 & 15.25 & 27.78 & 16.67 & 19.05 & 14.94 & 20.00 & 13.33 \\ 
        ~ & Overloading & 3.81 & 1.11 & 1.96 & 2.90 & 2.98 & 0.00 & 2.82 & 0.00 & 0.00 & 2.38 & 4.60 & 2.22 & 2.86 \\ 
        ~ & Bottleneck & 1.90 & 5.56 & 3.92 & 3.62 & 4.17 & 0.00 & 3.95 & 0.00 & 3.03 & 7.14 & 2.30 & 4.44 & 2.86  \\ 
          	
	Affective & Boring Task & 2.86 & 6.67 & 11.76 & 1.45 &  2.98 & 14.81 & 3.95 & 11.11 & 10.61 & 2.38 & 1.15 & 8.89 & 0.95   \\ 
        Factors & Lack of Motivation & 4.76 & 5.56 & 7.84 & 2.90 & 4.17 & 11.11 & 5.08 & 5.56 & 9.09 & 7.14 & 1.15 & 6.67 & 3.81  \\ 
	~ & Communication & 7.62 & 6.67 & 3.92 & 7.97 &  7.14 & 7.41 & 6.78 & 11.11 & 7.58 & 4.76 & 8.05 & 6.67 & 7.62  \\ 	
        ~ & Task Complexity & 11.43 & 8.89 & 11.76 & 10.14 & 10.12 & 11.11 & 10.73 & 5.56 & 9.09 & 11.90 & 10.34 & 13.33 & 7.62  \\        
        ~ & Noise Level & 0.00 & 2.22 & 1.96 & 0.72 &1.19 & 0.00 & 1.13 & 0.00 & 3.03 & 0.00 & 0.00 & 1.11 & 0.95  \\         
	Other & ~ & 0.95 & 3.33 & 0.00 & 2.90 & 2.38 & 0.00 & 2.26 & 0.00 & 1.52 & 0.00 & 3.45 & 4.44 & 0.00 \\ 
        \bottomrule
    \end{tabularx}
\end{table*}

\subsubsection{Stressors and Diversity}
We analyzed how the students experienced the different stressors, while taking the diversity dimensions from \Cref{tab:StudentDemographics} into account. \Cref{tab:heatmap-stressors-diversity} shows the relative number of mentioned stressors for each dimension of diversity. 
The results contained in the table are a mixed bag: some stressors, e.g., \emph{external disturbances}, \emph{missing work approach}, and \emph{time}, are more frequently mentioned than others. \Cref{tab:heatmap-stressors-diversity} also shows that these stressors are mentioned by the students across all diversity dimensions. However, a few stressors are noticeable, and we therefore analyze these in detail in the following. 



\begin{compactdesc}
	\item[Age:] The \emph{missing work approach} was the stressor most frequently mentioned by both age groups. \emph{Time management} was more of a concern for younger students, whereas \emph{time} as well as \emph{task distribution} ($p=0.028$) were challenging for older students.
	
	\item[Sex:] While \emph{time} and \emph{time management} were perceived particularly stressful for female students, male students were mainly stressed out by the \emph{missing work approach}. 
The difference is significant for \emph{time management} ($p=0.027$). Furthermore, females were more stressed by the \emph{boring task} ($p=0.002$) and by the \emph{lack of motivation} of team partners ($p=0.033$). One female student summarized this as follows: \emph{``Some team members were very distracted and did not take a decisive stand against this; everyone already had experience and wanted to push through their strategy. This led to starting difficulties and motivation decreases'' (UP1-12)}.

	\item[Ethnicity:] While local students were most stressed by a \emph{missing work approach}, non-local students were most affected by \emph{time} ($p=0.030$) as one non-local student explained: \emph{``Too little time in advance to discuss and to get a clear idea of what should be done and how to do it'' (UP2-1)}. 
Also, the perception regarding the \emph{boring task} was more stressful to non-local students ($p=0.016$).

	\item[Education:] The difference regarding the \emph{boring task} is also significant for different educational levels ($p=0.016$), i.e., students with Highschool degrees are less bored.
Noticeable, yet not statistically significant, is that students with \emph{other} educational backgrounds were more stressed by a lack of \emph{time}. Overall, most of the experiences of stress were similar for both groups.

	\item[Social Background:] We find some significant differences, particularly in the groups where both parents are academics compared to those without any academic background. 
In addition to a \emph{missing work approach}, which academic-students perceived more stressful ($p=0.012$), the stressors mentioned differed primarily in \emph{task distribution}: students with one ($p=0.016$) or both parents ($p=0.042$) as academics each felt it significantly more stressful than those without as one student with academical parents states: \emph{``We had to define a strategy; the roles in the team were not clear; these issues resulted in an unstructured teamwork.'' (UP2-12)}. 
Similarly, there are significant differences between both parents and none being academics regarding the affective factors, \emph{boring task} ($p=0.021$) and \emph{lack of motivation} ($p=0.048$).

	\item[Work Experience:] Students with more than half a year of work experience perceive the \emph{boring task} as significantly more stressful than those with less work experience ($p=0.006$). 
Furthermore, those with work experience perceived \emph{time} ($p=0.042$) and \emph{external disturbances} ($p=0.038$) as significantly greater stressors than others. 

\end{compactdesc}

\rqsummary{RQ1}{The students considered \emph{missing work approach, time and time management, communication}, and \emph{external disturbances} as main stressors. The diversity dimension of \emph{social background} impacts the perceived stressors the most. There are significant differences in five out of six diversity dimensions of the stressor \emph{boring task}.}
\rqinterpretation{RQ1}{Our results show that a missing work approach, time, and organization issues challenge students the most. Even though the perception of stress changes over time, we assume that guidance on self-organizing teams for student projects may be helpful, since certain diversity groups seem to react much more strongly to certain stressors. Since such stressors are common in professional software development, students should also be provided with effective strategies to overcome or address them, e.g., by establishing a communication routine.}



\subsection{RQ2: Perceived Learnings and Diversity}
\label{sec:Results:RQ2}
\subsubsection{Extraction and Analysis of Perceived Learnings}
Based on the post-experiment questionnaire, we analyzed 322 mentions in total 
and extracted 17 perceived learnings. \Cref{tab:heatmap-learnings-runs} shows the absolute number of perceived learnings for each sub-experiment across all three experiments. 
Again, we used the top-level categories \emph{organizational factors} (47 mentions), \emph{planning \& strategy} (202), \emph{affective factors} (67), and \emph{other}~(6).

\begin{table}[tb]
	\centering
	\vspace{-0.5em}
	\caption{Mentioned learnings per sub-experiment.}
	\vspace{-0.5em}
	\label{tab:heatmap-learnings-runs}
	\begin{tabularx}{\linewidth}{lXCCCr}
	\toprule
        Category & Subcategory &  \multicolumn{1}{c}{HHZ} & \multicolumn{1}{c}{UP1} & \multicolumn{1}{c}{UP2} & \multicolumn{1}{c}{$\sum$} \\ 
	\midrule
	Organisat. & Working Alone & 2 & 1 & 0 & 3 \\ 
         Factors & Smaller Teams' Effec. & 5 & 5 & 0 & 10 \\ 
         & Bigger Teams' Effec. & 3 & 6 & 7 & 16 \\ 
	& Knowledge Growth & 1 & 8 & 6 & 15 \\ 	
         & Quiet Space & 0 & 0 & 3 & 3 \\ 
	Planning & Established Work Appr. & 14 & 36 & 25 & 75 \\ 
          \& Strategy & Developed Routine & 0 & 8 & 7 & 15 \\ 
         & Work Efficiency & 6 & 4 & 4 & 14 \\  
         & Improved Adaptability & 0 & 1 & 8 & 9 \\ 	    
         & Effective Task Distrib. & 14 & 18 & 15 & 47 \\ 
	& Established Time Manag. & 0 & 6 & 5 & 11 \\
	 & Relaxed Schedule  & 0 & 11 & 13 & 24 \\  
	& Appointed Recorder & 1 & 3 & 3 & 7 \\      
	Affective  & Communication & 11 & 7 & 9 & 27 \\ 
         Factors& Motivation  & 15 & 4 & 4 & 23 \\ 
         & Satisfaction & 7 & 4 & 3 & 14 \\ 
	& Creativity & 3 & 0 & 0 & 3 \\ 
        Other &  & 2 & 0 & 4 & 6 \\ 
        $\sum$ &  & 84 & 122 & 116 & 322 \\ 
	\bottomrule
	\end{tabularx}
\end{table}

The analysis of the qualitative data using the 17 categories shows that developing an \emph{established work approach} (75) and developing an approach for an \emph{effective task distribution}~(47) were considered the most important subjective learnings, followed by \emph{communication} and \emph{relaxed schedule}.
\begin{figure}[!t]
	\centering
	\includegraphics[width=\linewidth]{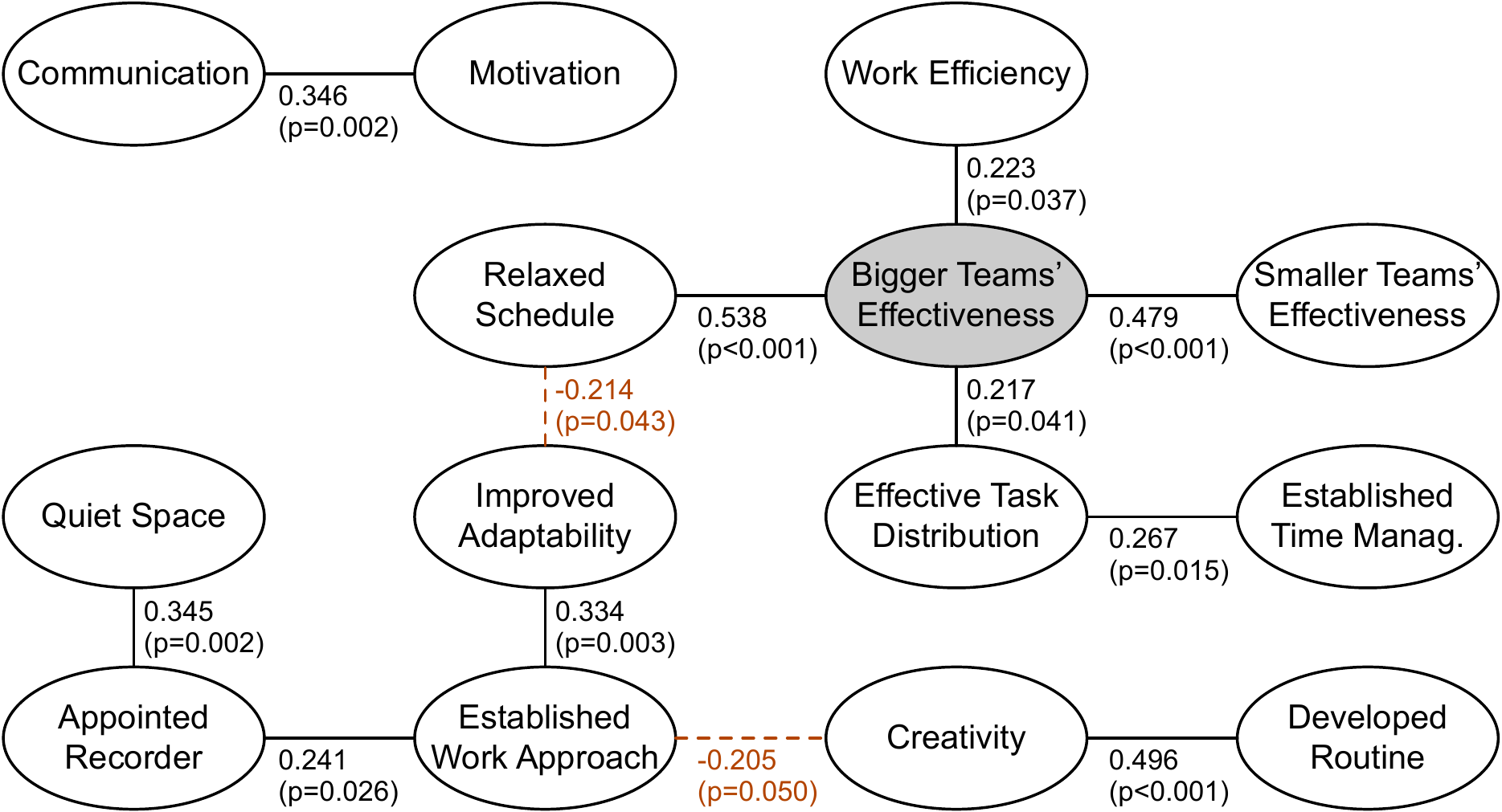}
	\caption{Linear relations between mentioned learnings.}
	\label{fig:Results:RQ2:LinearLearnings}
\end{figure}
\begin{table*}[tb]
	\centering
	\vspace{-0.5em}
	\caption{Percentage of mentioned learnings from the students belonging to a certain dimension of diversity.}
	\vspace{-0.5em}
	\label{tab:heatmap-learnings-diversity}
	 \begin{tabularx}{\linewidth}{XlCCCCCCCCCCCCCC}
    	\toprule
        Category & Subcategory & \multicolumn{2}{c}{Age}  & \multicolumn{2}{c}{Sex}  & \multicolumn{2}{c}{Ethnicity}  & \multicolumn{2}{c}{Education}& \multicolumn{3}{c}{Social BG} & \multicolumn{2}{c}{Work} \\ 
        ~ & ~ & \multicolumn{1}{c}{$<$21} & \multicolumn{1}{c}{$\geq$21} & \multicolumn{1}{c}{f} & \multicolumn{1}{c}{m} & \multicolumn{1}{c}{loc.} & \multicolumn{1}{c}{oth.}& \multicolumn{1}{c}{high.} & \multicolumn{1}{c}{oth.} & \multicolumn{1}{c}{both} & \multicolumn{1}{c}{one} & \multicolumn{1}{c}{none} & \multicolumn{1}{c}{$\geq$0.5y} & \multicolumn{1}{c}{$<$0.5y} \\ 
        \midrule
        Organizat. & Working Alone & 0.00 & 3.33 & 1.96 & 1.45 &  1.19 & 3.70 & 1.69 & 0.00 & 3.03 & 0.00 & 1.15 & 1.11 & 1.90  \\ 
        Factors & Smaller Teams' Eff. & 7.62 & 2.22 & 3.92 & 5.07 &  5.36 & 3.70 & 5.08 & 5.56 & 4.55 & 4.76 & 5.75 & 1.11 & 8.57  \\ 
        ~ & Bigger Teams' Eff. & 9.52 & 6.67 & 9.80 & 7.25 &  8.93 & 3.70 & 7.91 & 11.11 & 9.09 & 4.76 & 9.20 & 10.00 & 6.67 \\ 
        ~ & Knowledge Growth & 4.76 & 11.11 & 11.76 & 5.07 &  6.55 & 14.81 & 7.34 & 11.11 & 9.09 & 4.76 & 8.05 & 10.00 & 5.71  \\ 
	~ & Quiet Space & 0.95 & 2.22 & 0.00 & 2.17 & 1.79 & 0.00 & 1.69 & 0.00 & 3.03 & 0.00 & 1.15 & 3.33 & 0.00 \\ 
     
	Planning, & Estab. Work Appr. & 38.10 & 38.89 & 39.22 & 37.68 &  36.90 & 48.15 & 37.85 & 44.44 & 40.91 & 42.86 & 34.48 & 36.67 & 40.00  \\       
        Strategy & Developed Routine & 6.67 & 8.89 & 7.84 & 7.97 &  7.14 & 11.11 & 8.47 & 0.00 & 9.09 & 2.38 & 9.20 & 8.89 & 6.67\\ 
        ~ & Work Efficiency & 4.76 & 10.00 & 7.84 & 6.52 &  7.74 & 3.70 & 4.52 & 33.33 & 4.55 & 9.52 & 8.05 & 7.78 & 6.67 \\ 
        ~ & Improved Adaptability & 2.86 & 6.67 & 3.92 & 5.07 &  4.17 & 7.41 & 3.95 & 11.11 & 3.03 & 4.76 & 5.75 & 4.44 & 4.76 \\ 
        ~ & Effective Task Distrib. & 21.90 & 26.67 & 23.53 & 24.64 &  26.79 & 7.41 & 22.03 & 44.44 & 21.21 & 28.57 & 24.14 & 27.78 & 20.95  \\ 
        ~ & Estab. Time Manag. & 7.62 & 3.33 & 5.88 & 4.35 &  6.55 & 0.00 & 5.65 & 5.56 & 6.06 & 2.38 & 6.90 & 7.78 & 3.81  \\  
	~ & Relaxed Schedule & 7.62 & 17.78 & 15.69 & 11.59 &  11.31 & 18.52 & 12.99 & 5.56 & 9.09 & 11.90 & 14.94 & 18.89 & 6.67  \\ 
	~ & Appointed Recorder & 1.90 & 5.56 & 3.92 & 3.62 &  3.57 & 3.70 & 3.95 & 0.00 & 4.55 & 4.76 & 2.30 & 3.33 & 3.81  \\ 
	        
        Affective & Communication & 18.10 & 8.89 & 9.80 & 14.49 & 13.69 & 14.81 & 13.56 & 16.67 & 12.12 & 14.29 & 14.94 & 15.56 & 12.38  \\ 
        Factors & Motivation & 11.43 & 12.22 & 17.65 & 10.14 &  10.71 & 18.52 & 10.17 & 27.78 & 12.12 & 14.29 & 10.34 & 16.67 & 7.62  \\ 
        ~ & Satisfaction & 8.57 & 5.56 & 3.92 & 8.70 &  7.14 & 7.41 & 7.91 & 0.00 & 3.03 & 7.14 & 10.34 & 5.56 & 8.57   \\ 
        ~ & Creativity & 0.95 & 2.22 & 1.96 & 1.45 &  1.79 & 0.00 & 1.69 & 0.00 & 3.03 & 0.00 & 1.15 & 2.22 & 0.95  \\ 
  
        Other & ~ & 3.81 & 2.22 & 1.96 & 3.62 & 2.98 & 3.70 & 3.39 & 0.00 & 1.52 & 4.76 & 3.45 & 4.44 & 1.90 \\ 
        \bottomrule
    \end{tabularx}
\end{table*}

\subsubsection{Relationships between Perceived Learnings}
\Cref{fig:Results:RQ2:LinearLearnings} shows the significant linear relations between perceived learnings. Similar to the stressors, not all mentioned learnings (14 out of the 17 categories) have significant relations. The majority of perceived learnings is connected to \emph{bigger teams' effectiveness}, which forms a center that connects \emph{work efficiency} and \emph{effective task distribution}---learnings that can be connected to the stressors from \Cref{fig:Results:RQ1:LinearStressors}. The participants consider an \emph{established work approach} limiting \emph{creativity}, and a \emph{relaxed schedule} negatively relating to \emph{improved adaptability}.

\subsubsection{Perceived Learnings and Diversity}
Based on analyzed perceived learnings, we investigate how the participants experience the learnings taking the different diversity dimensions from~\Cref{tab:StudentDemographics} into account. \Cref{tab:heatmap-learnings-diversity} shows the relative number of mentioned stressors in the questionnaire for each dimension of diversity. The table shows that \emph{established work approach} is a perceived learning across all diversity dimensions, as well as the \emph{effective task distribution}. However, there also exist differences, which we explain in more detail in the following.



\begin{compactdesc}
	\item[Age:] \emph{Knowledge growth} as well as \emph{relaxed schedule} due to more time was mentioned significantly more often by older students than by younger ones ($p=0.036$). 
A strong learning effect was experienced by younger students on \emph{communication}, although this is not significant.

	\item[Sex:] Female students perceive \emph{knowledge growth} more than male students ($p=0.028$). Increasing \emph{motivation} was also stronger for women, although not significantly so. Overall, perceived learnings are similar between sexes.

	\item[Ethnicity:] While non-local students were significantly more likely to report a \emph{knowledge growth} as a learning ($p=0.033$), for local students, it is the \emph{effective task distribution} ($p=0.036$). One non-local student stated: \emph{``The process went well. All the teammates had the same strategy. When new guys came in [sic!] they were well instructed--- the process was not interrupted'' (UP2-5)}
	Besides these two, the other perceived learnings are similar for both groups of this dimension.

	\item[Education:] \emph{Work efficiency} ($p=0.004$) and \emph{motivation} ($p=0.012$) are significantly more commonly perceived learnings for students with other degrees than for those with high school degrees. Similarly, students in the \emph{education: other} category mentioned the learning of \emph{effective work distribution} twice as often, although this is not statistically significant. One student with another degree reported: ``\emph{Agreements and strategies were implemented immediately. Suggestions for improvements were accepted and implemented; everything seemed as if we had always worked as a team''~(UP2-1)}.
	
	\item[Social Background:] There are no significant differences in the perceived learnings for the social background dimension, suggesting that learnings for this dimension are universal.
	
	\item[Work Experience:] Team member \emph{motivation} was experienced by more students who already had work experience as learning ($p=0.016$). Similarly, they valued the \emph{relaxed schedule} due to more available time as higher learning ($p=0.027$). There are also significant differences in the organizational factors of \emph{smaller teams efficency} ($p=0.006$) and \emph{quiet space} ($p=0.028$).

\end{compactdesc}

\rqsummary{RQ2}{The students considered \emph{establishing a work approach} and developing an \emph{effective task-distribution strategy} the most important perceived learnings. The diversity dimensions \emph{age} and \emph{working experience} influence the subjective learning experience the most. There were significant differences in knowledge growth for half of the diversity dimensions.}
\rqinterpretation{RQ2}{We can see that certain perceived learnings are connected to the stressors, e.g., work approach. We assume that there is value in balancing the stressors to emphasize desired learning outcomes. Certain team configurations with respect to diversity can support such a balancing. Here, it might be relevant to counteract potential stereotypes of age and experience of SE team members when dealing with technical challenges such as innovative or best practice development methods and technologies.}



\subsection{RQ3: Experiences of Diverse SE Teams}
\label{sec:Results:RQ3}

\subsubsection{Computing the Blau Index}
Having analyzed stressors and learnings at the individual level, we now focus on the teams. As described in \Cref{sec:methodology}, 11 teams of size five were selected for which we computed the \emph{Blau} index for each diversity dimension (\Cref{tab:TeamDiversityIndexBlau}). The diversity scores show that the teams range from homogeneous, e.g., teams in which all participants have the same education background, to fairly diverse teams, e.g., regarding social background. 

\begin{table}[!t]
\renewcommand{\arraystretch}{0.95}
\footnotesize
\vspace{-0.5em}
\caption{Teams diversity using the \emph{Blau} index.}
\vspace{-0.5em}
\label{tab:TeamDiversityIndexBlau}
	\begin{tabularx}{\linewidth}{Xrrrrrr}
	\toprule
	Team & $\text{B}_{\emph{age}}$ & $\text{B}_{\emph{sex}}$ & $\text{B}_{\emph{eth}}$ 
					& $\text{B}_{\emph{edu}}$ & $\text{B}_{\emph{soc}}$ & $\text{B}_{\emph{wrk}}$ \\
	\midrule
	HHZ-3-A     & 0.48 & 0.56 & 0.32 & 0.48 & 0.00 & 0.32 \\
	HHZ-3-B     & 0.32 & 0.56 & 0.32 & 0.00 & 0.56 & 0.48 \\
	HHZ-3-C     & 0.32 & 0.48 & 0.00 & 0.00 & 0.00 & 0.48 \\
	HHZ-3-D (c) & 0.48 & 0.32 & 0.32 & 0.00 & 0.48 & 0.32 \\
	\midrule
	UP1-3-A	    & 0.00 & 0.48 & 0.00 & 0.00 & 0.32 & 0.00 \\
	UP1-3-B     & 0.48 & 0.48 & 0.00 & 0.00 & 0.48 & 0.48 \\
	UP1-3-C (c) & 0.32 & 0.00 & 0.32 & 0.48 & 0.64 & 0.48 \\
	\midrule
	UP2-3-A     & 0.48 & 0.32 & 0.48 & 0.32 & 0.64 & 0.48 \\
	UP2-3-B     & 0.48 & 0.00 & 0.32 & 0.00 & 0.64 & 0.48 \\
	UP2-3-C     & 0.32 & 0.00 & 0.00 & 0.00 & 0.48 & 0.32 \\
	UP2-3-D (c) & 0.48 & 0.00 & 0.32 & 0.32 & 0.32 & 0.48 \\	
	\bottomrule
\end{tabularx}
\end{table}


\subsubsection{Stressors and Learnings of the Teams}
On average, teams identified eight stressors in the two runs of interest and, likewise, 10 learnings~(\Cref{tab:TeamLearning}). The most frequently mentioned stressors were \emph{external disturbances} (24 mentions), \emph{lack of motivation} (9), and \emph{boring task} (8), while the most frequently mentioned learnings were \emph{established work approach} (23), \emph{effective task distribution} (14), and \emph{relaxed schedule} (13).

\begin{table}[!t]
\renewcommand{\arraystretch}{0.95}
\footnotesize
\vspace{-0.5em}
\caption{Mentioned stressors and learnings per team \\ \small{(\emph{O}: orga. f., \emph{P}: planning and strategy, \emph{A}: affective f., \emph{M}: misc).}}
\vspace{-0.5em}
\label{tab:TeamLearning}
	\begin{tabularx}{\linewidth}{XRRRRrRRRRr}
	\toprule
	Team & \multicolumn{5}{c}{Identified Stressors} & \multicolumn{5}{c}{Perceived Learnings} \\
	& \multicolumn{1}{c}{\emph{O}} &  \multicolumn{1}{c}{\emph{P}} &  \multicolumn{1}{c}{\emph{A}} &  \multicolumn{1}{c}{\emph{M}} & \emph{$\Sigma$}  
	& \multicolumn{1}{c}{\emph{O}} &  \multicolumn{1}{c}{\emph{P}} &  \multicolumn{1}{c}{\emph{A}} &  \multicolumn{1}{c}{\emph{M}} & \emph{$\Sigma$}  \\
	\midrule
	HHZ-3-A     & 5 & 2 &  0 &  0&  7 &  9 & 2 & 1 &  0  & 12 \\
	HHZ-3-B     & 1 & 1 & 5 & 0  &  7 &  3 & 2 & 5 &  0  & 10 \\
	HHZ-3-C     & 6 & 3 & 1 & 0  & 10 &  7 & 2 & 3 &  0  & 12 \\
	HHZ-3-D (c) & 0  &  0 & 3 & 1 &  4 & 10 & 2 & 0  &  0  & 12 \\
	\midrule
	UP1-3-A	   & 5 & 4 & 2 & 2 & 13 &  3 & 2 & 1 & 1 &  7 \\
	UP1-3-B     & 5 & 3 & 3 & 0  & 11 &  4 & 1 &  0 & 1 &  6 \\
	UP1-3-C (c) & 2 & 3 & 5 &  0 & 10 &  7 & 3 &  0 &  0 & 10 \\
	\midrule
	UP2-3-A     & 5 &  0 & 1 &  0 &  6 & 10 & 3 &  0 & 0  & 13 \\
	UP2-3-B     & 5 & 1 & 1 &  0 &  7 &  5 & 2 & 2 & 2 & 11 \\
	UP2-3-C     & 7 & 1 & 0  &   0&  8 &  4 & 2 & 3 &  0 &  9 \\
	UP2-3-D (c) & 1 & 7 & 2 &  0& 10 &  6 & 3 & 1 &  0 & 10 \\	
	\bottomrule
\end{tabularx}
\end{table}

%

\begin{table}[!t]
\renewcommand{\arraystretch}{0.95}
\footnotesize
\vspace{-0.5em}
\caption{Teams diversity, stressors and learnings.}
\vspace{-0.5em}
\label{tab:TeamDiversityVsStressAndLearning}
	\begin{tabularx}{\linewidth}{Xrrrrrr}
	\toprule
	Dimension & \multicolumn{2}{c}{Stressors} & \multicolumn{2}{c}{Learnings} \\
	\textbf{} & $r$ & $p$-value & $r$ & $p$-value \\
	\midrule
	$\text{B}_{\emph{age}}$ &\cellcolor{gray!70} -0.6205 & \cellcolor{gray!70}0.041 &  \cellcolor{gray!50}0.4596 & \cellcolor{white}0.155 \\
	$\text{B}_{\emph{sex}}$ & \cellcolor{white}0.0486 & \cellcolor{white}0.887 & \cellcolor{white}-0.0493 & \cellcolor{white}0.885 \\
	$\text{B}_{\emph{eth}}$ &\cellcolor{gray!70} -0.6504 & \cellcolor{gray!70}0.030 &  \cellcolor{gray!70}0.6657 & \cellcolor{gray!70}0.025 \\
	$\text{B}_{\emph{edu}}$ & \cellcolor{white}-0.0460 & \cellcolor{white}0.893 &  \cellcolor{gray!50}0.3588 &\cellcolor{white} 0.278 \\
	$\text{B}_{\emph{soc}}$ & \cellcolor{white}-0.2501 & \cellcolor{white}0.458 & \cellcolor{white}-0.1309 & \cellcolor{white}0.701 \\
	$\text{B}_{\emph{wrk}}$ & \cellcolor{gray!50} -0.3014 & \cellcolor{white}0.367 &  \cellcolor{gray!50}0.3478 & \cellcolor{white}0.294 \\
	\bottomrule
\end{tabularx}
\end{table}

We cannot find a statistically significant relationship between a team's diversity and both stressors or perceived learnings at the top level~(\cref{tab:TeamDiversityVsStressAndLearning}). 
However, at the individual level, we find several (significant) relationships. 
\begin{compactdesc}
	\item[Age:] The \emph{age} is negatively correlated with the stressors ($r=-0.6205,  p=0.041$), indicating that teams with greater age diversity experience fewer stressors. Although not significant, there is a moderate correlation between age and perceived learning~($r=0.4596,  p=0.155$).

	\item[Ethnicity:] The \emph{ethnicity} is negatively correlated with the stres\-sors ($r=-0.6504,  p=0.030$), and positively correlated with the perceived learnings~($r=0.6657,  p=0.025$). This implies that teams with higher ethnic diversity experience fewer stressors, yet have more learnings.

	\item[Work Experience:] The \emph{work experience} has medium, but not significant, correlations with stressors ($r=-0.3014,  p=0.367$) and perceived learnings ($r=0.3478, p=0.294$).
\end{compactdesc}
When considering both teams with the highest (UP2-3-A) and lowest (UP1-3-A) overall diversity~(\Cref{tab:TeamDiversityIndexBlau}), we see a tendency for the most diverse team to report less than half the stressors and almost twice as many learnings as the most homogeneous team (\Cref{tab:TeamLearning}). The most heterogeneous team also did not indicate any stressors in \emph{planning \& strategy}.

Nevertheless, Table~\ref{tab:TeamDiversityVsStressAndLearning} shows that, even though a few significant correlations exist, $\text{H1}_{0}$ and $\text{H2}_{0}$ cannot be rejected as the \emph{Bonferroni}-corrected $p_{\text{B\_cor}}\leq0.0083$ is not reached.

\rqsummary{RQ3}{Heterogeneous teams, especially those with a variety of \emph{ages, ethnicities}, and \emph{work experiences}, tend to experience fewer stressors, but perceive more learnings than homogeneous teams. However, $\text{H1}_{0}$ and $\text{H2}_{0}$ cannot be rejected, i.e., we still have to assume that there is no relation between diversity and stressors and perceived learnings.}
\rqinterpretation{RQ3}{Our results indicate some influence of diversity on perceived stress and perceived learnings. We assume that respecting diversity during team formation may have an impact on the team's stress tolerance and the learning opportunities and, eventually, might affect the team's overall performance, notably, to focus on the actual development tasks even when working under pressure.}


\subsection{RQ4: Effects on SE Team Performance}
\label{sec:Results:RQ3}
\subsubsection{Computing the Performance Scores}
Based on the diversity scores (\Cref{sec:Results:RQ3}, \Cref{tab:TeamDiversityIndexBlau}), we study whether diversity in teams relates to team performance and work quality.
\Cref{tab:TeamPerformance} shows the numbers used for the analyses. 

\subsubsection{Performance of the Teams}
As \Cref{tab:TeamDiversityVsPerformance} shows, there are no statistically significant relationships between the diversity dimensions and the counting performance and quality of work. Considering the counting performance, there is a medium relationship between the diversity dimension \emph{education} and the average performance. However, the analysis of the diversity dimensions and their relationship to the quality of work reveals that only the diversity dimension \emph{social background} shows a weak (no) relationship. 
All other dimensions have a medium, yet not significant, relationship with the quality of work. It can also be seen that there is no large deviation in the clerical error rates of the control groups. \Cref{tab:TeamPerformance} shows that the control groups usually have a smaller relative error, yet, the clerical error does not follow the trend. 
It must also be noted that team \emph{B} from the third experiment (UP) has a 100\% relative error with a clerical error rate of 0\%. This means that this group did not provide any controlling data under time pressure.

\begin{table}[!t]
\renewcommand{\arraystretch}{0.95}
\footnotesize
\vspace{-0.5em}
\caption{Team performance numbers.}
\vspace{-0.5em}
\label{tab:TeamPerformance}
	\begin{tabularx}{\linewidth}{Xrrrrrr}
	\toprule
	Team & Counted Items & Relative Error & Clerical Error \\
	\midrule
	HHZ-3-A     & 747.50 & 78.49\% & 13.30\% \\
	HHZ-3-B     & 478.50 & 25.44\% & 60.00\% \\
	HHZ-3-C     & 692.00 & 47.08\% & 43.33\% \\
	HHZ-3-D (c) & 398.50 & 12.14\% & 42.22\% \\
	\midrule
	UP1-3-A	    & 539.50 & 53.48\% & 76.98\% \\
	UP1-3-B     & 442.00 & 49.41\% & 84.44\% \\
	UP1-3-C (c) & 774.50 & 37.10\% & 20.59\% \\
	\midrule
	UP2-3-A     & 521.50 & 10.62\% & 46.22\% \\
	UP2-3-B     & 858.00 & 100\%   &  0.00\% \\
	UP2-3-C     & 311.00 & 22.96\% & 44.09\% \\
	UP2-3-D (c) & 387.50 & 15.35\% & 30.56\% \\	
	\bottomrule
\end{tabularx}
\end{table}

\begin{table}[!t]
\renewcommand{\arraystretch}{0.95}
\footnotesize
\vspace{-0.5em}
\caption{Teams diversity, performance \& quality indicators.}
\vspace{-0.5em}
\label{tab:TeamDiversityVsPerformance}
	\begin{tabularx}{\linewidth}{Xrrrrrrcc}
	\toprule
	Dimension & \multicolumn{2}{c}{Counted Items} & \multicolumn{2}{c}{Clerical Error} \\
	~ & $r$ & $p$-value & $r$ & $p$-value \\
	\midrule
	$\text{B}_{\emph{age}}$ & \cellcolor{white} 0.0236 & \cellcolor{white}0.945 & \cellcolor{gray!50}-0.4443 & \cellcolor{white}0.170 \\
	$\text{B}_{\emph{sex}}$  & \cellcolor{white}-0.0113 & \cellcolor{white}0.973 & \cellcolor{gray!50}0.5350 & \cellcolor{white}0.089 \\
	$\text{B}_{\emph{eth}}$  & \cellcolor{white} 0.2301 & \cellcolor{white}0.496 &\cellcolor{gray!50}-0.5500 & \cellcolor{white}0.079 \\
	$\text{B}_{\emph{edu}}$ &  \cellcolor{gray!50}0.3367 & \cellcolor{white}0.311 & \cellcolor{gray!50}-0.4943 & \cellcolor{white}0.122 \\
	$\text{B}_{\emph{soc}}$  & \cellcolor{white}-0.1269 & \cellcolor{white}0.710 &  \cellcolor{white}0.0218 & \cellcolor{white}0.949 \\
	$\text{B}_{\emph{wrk}}$ &  \cellcolor{white}0.1652 & \cellcolor{white}0.627 & \cellcolor{gray!50}-0.3293 &\cellcolor{white} 0.322 \\
	\bottomrule
\end{tabularx}
\end{table}

In a nutshell, \emph{Pearson's r} does not yield a significant relationship between the diversity dimensions and the performance. As shown by Table~\ref{tab:TeamDiversityVsPerformance}, this means that $\text{H3}_{0}$ cannot be rejected as the \emph{Bonferroni}-corrected $p_{\text{B\_cor}}\leq0.0042$ is not reached.
However, we identify medium correlations, notably, between five out of the six diversity dimensions and the quality of work. From these five medium relationships, four are negatively correlated, i.e., 
as the variety of \textit{age, ethnicity, educational background} and \textit{work experience} increases, the rate of clerical error decreases, thus, the quality of the work increases. 
Furthermore, there is a medium positive correlation between the diversity dimension \emph{education background} and the number of items counted ($r=0.3367, p=0.311$). That is, with the diversity of education in a team, the number of counted items increases while the error rate decreases (increased quality of work) ($r=-0.4943, p=0.122$), which leads to a higher performance of such teams. 

Considering the teams with the highest (UP2-3-A) and lowest (UP1-3-A) diversity~(\Cref{tab:TeamDiversityIndexBlau}), we notice that for a comparable number of counted items, the relative error is only one fifth of that of the most homogeneous team and, in fact, is the lowest overall error rate. Similarly, the most diverse team has 30\% less clerical errors than the most homogeneous team.

\rqsummary{RQ4}{We find no significant relationships between diversity of teams and performance. Even though there is a medium correlation between the quality of work and five out of six diversity dimensions, we cannot reject $\text{H3}_{0}$. That is, we still have to assume that there is no relation between diversity and team performance.}
\rqinterpretation{RQ4}{Despite the absence of significant relationships we assume that respecting diversity during team formation might affect overall performance: Our results indicate that increased diversity might have an effect on the quality of work. Given that software has become crucial to almost every aspect of the society, every chance to improve the quality of software must be taken. Errors not made in the initial development do not cause trouble and do not need to be expensively fixed during a software's operation phase.}


\section{Conclusions and Future Work}
\label{conclusions}

In this paper, we exposed 65 SE students in a series of three experiments to stress in teamwork to prepare them for their professional future. We analyzed the identified 19 stressors and 17 positive perceived learnings depending on six dimensions of diversity for the individual students as well as for the teams.

\paragraph*{Summary of Findings}
Our findings include that students with different \emph{social background} show the most differences in perceiving stressors, whereas this diversity dimension has no influence on the perceived learnings. For the perceived learnings, \emph{age} and \emph{work experience} are the most influential diversity dimensions. At the team level we observed significant relations for the dimensions \emph{age} and \emph{ethnicity}. Setting up teams with respect to these diversity dimensions might have a positive effect on the team's health, as an increased diversity might reduce stress and foster subjective learning. 
An intriguing observation is that the dimension \emph{sex}, even though it only has a small influence in general, affected stress more than perceived learning. In this regard, notably the most diverse team consisted of four female students (\Cref{fig:experiment}), who also had a completely different work approach than all other teams (team members were standing around the table).

\paragraph*{Takeaways for Teachers}
Our research has some takeaways for teachers, which could help them improve their project courses. A key contribution of this paper is the analysis of relationships among stressors (\Cref{fig:Results:RQ1:LinearStressors}) and, likewise, perceived learnings (\Cref{fig:Results:RQ2:LinearLearnings}). These results provide means to form teams by providing them with better starting conditions. For example, providing a clear organization and a well-defined work approach is likely to have a positive impact. Especially in times of agile software development, this recommendation may be surprising, as a flexible work approach and self-organizing teams are key components of agile software development. However, this does not seem to work as well for student teams. Furthermore, our findings can help teachers analyze dysfunctional teams, i.e., to analyze why a team is in trouble, e.g., whether the task is not clearly described, or whether the task distribution is inappropriate. 

Finally, our findings provide some help for better balancing teams: Some diversity dimensions, e.g., \emph{social background}, have a relationship to stressors and perceived learnings that allow for setting up teams that are more stress-resistant. Since this sensitive information is difficult to obtain in a regular classroom setting, raising awareness among educators about various backgrounds of students and a deeper understanding of the root causes of high or low performance provide a baseline. Such insights about different diversity dimensions can be used to improve the alignment of team work for project assignments, perceived learning objectives of a course and interventions within the teams.

\paragraph*{Human Factors in Software Engineering}
The importance of human factors in SE is still far from being a major focus in many SE education programs. For instance, even though the ACM/IEEE curriculum design guideline states \emph{``Although, for the purposes of curriculum design, these are not subject areas needing deep study [...]''}, still, the guideline continues as follows: \emph{``software engineers must be aware of the effects that human factors can have across many of the discipline's activities''}~\cite[p.~17]{ACMCurricula:2015aa}. It is stated that \emph{``Students need to repeatedly see how software engineering is not just about technology''}~\cite[p.~40]{ACMCurricula:2015aa}, however, basic human factors in the computing essentials topics~\cite[CMP.cf.6]{ACMCurricula:2015aa} mostly refer to I/O, errors, robustness, or software testing involving users.
Therefore, it is key to emphasize social skills such as communication and collaboration abilities in introduction lectures on SE and project work. Students need to be taught that social skills, human factors, and technical skills are equally relevant, especially in the context of agile software development, which has to be considered a \emph{cultural topic} of project teams and software-producing organizations~\cite{9496156}.

\paragraph*{Future Work}
To our surprise, we found no significant influence of diversity on team performance and quality of work. However, even though not significant, the diversity dimension \emph{social background} has a medium relationship to the quality of work, which motivates further studies. For this, planned future work includes repeating this experiment to improve our data and to study those aspects with medium relationships in more detail. It would also be beneficial to replicate this experiment at other universities, and even in companies, to acquire more insights on whether stress and learning perceptions change over time. Our study was based on the intentionally simple task of sorting and counting sweets to minimize potential conflating factors; given our results as a baseline, further replications could now aim to generalize our findings by considering more complex software engineering tasks.
As we provide a replication package, we cordially invite other teachers and researchers to use the package and to independently replicate our study.

\section*{Acknowledgements}\label{sec:acknowledgements}
This work is supported by the Federal Ministry of Education and Research
through project ``primary::programming'' (01JA2021) as
part of the ``Qualitätsoffensive Lehrerbildung'', a joint initiative of the
Federal Government and the Länder. The authors are responsible for the content
of this publication.

\bibliographystyle{IEEEtran}
\bibliography{references}

\end{document}